\documentclass{ws-procs9x6}
\usepackage[latin1]{inputenc}
\begin{document}

\title{A low background Micromegas detector for the CAST
experiment}

\author{\uppercase{P. Abbon, S. Andriamonje, S. Aune, D. Besin,  S. Cazaux,
P.~Contrepois, N. Duportail, E. Ferrer Ribas, M. Gros, I.~G.~
Irastorza, A. Giganon, I. Giomataris, M. Riallot and
G.~Zaffanela}}
\address{DAPNIA, Centre d'études de Saclay, 91191 Gif sur Yvette CEDEX, France}

\author{G. Fanourakis, T. Geralis, K. Kousouris and K. Zachariadou}
\address{Institute of Nuclear Physics, NCSR Demokritos, Aghia Paraskevi 15310, Athens, Greece}

\author{T. DAFNI\footnote{Now at \uppercase{DAPNIA}, \uppercase{C}entre d'\'{e}tudes de \uppercase{S}aclay, 91191 \uppercase{G}if sur \uppercase{Y}vette \uppercase{CEDEX}, \uppercase{F}rance}}
\address{Institut f\"{u}r Kernphysik, TU-Darmstadt, Schlossgartenstr. 9, D-64289 Darmstadt, Germany}
\author{\uppercase{T. Decker, R. Hill, M. Pivovaroff and R. Soufli}}
\address{Lawrence Livermore National Laboratory, 7000 East Avenue,
Livermore CA94550, USA}

\author{J. MORALES}
\address{ Instituto de F\'{\i}sica Nuclear y Altas Energ\'{\i}as, Universidad
de Zaragoza, Zaragoza 50009, Spain}


\maketitle

\abstracts{A low background Micromegas detector has been operating
on the CAST experiment at CERN for the search of solar axions
during the first phase of the experiment (2002-2004). The detector
operated efficiently and achieved a very low level of background
rejection ($5\times 10^{-5}$ counts keV$^{-1}$cm$^{-2}$s$^{-1}$)
thanks to its good spatial and energy resolution as well as the
low radioactivity materials used in the construction of the
detector. For the second phase of the experiment (2005-2007), the
detector will be upgraded by adding a shielding and including
focusing optics. These improvements should allow for a background
rejection better than two orders of magnitude.}

\section{Introduction}
The CAST (Cern Axion Solar Telescope) collaboration is using a
decommissioned LHC dipole magnet to convert solar axions into
detectable x-ray photons. Axions are light pseudoscalar particles
that arise in the context of the Peccei-Quinn\cite{PecceiQuinn}
solution to the strong CP problem and can be Dark Matter
candidates\cite{Sikivie}. Stars could produce axions via the
Primakoff conversion of the plasma photons. The CAST experiment
aims to track the Sun in order to detect solar axions. The
detection principle is based on the coupling of an incoming axion
to a virtual photon provided by the transverse field of an intense
dipole magnet, being transformed into a real, detectable photon
that carries the energy and the momentum of the original axion.
The axion to photon conversion probability is proportional to the
square of the transverse field of the magnet and to the active
length of the magnet. Using an LHC magnet (9~T and 9.26~m long)
improves the sensitivity by a factor 100 compared to previous
experiments. A more detailed description of the principle of the
experiment can be found in Zioutas {\it et al.}\cite{Zioutas}.

For the first phase of the experiment, 2002-2004, three different
types of detectors have been developed to detect the x-rays
originated by the conversion of the axions inside the vacuum of
the magnet:~a time projection chamber (TPC), a CCD and a
Micromegas detector. The CCD detector has been working in
conjunction with a mirror system to focus x-rays coming out of the
magnet bores improving its signal to background ratio. The
analysis of the data collected during 2003 shows no excess of
signal over background and has resulted in the most restrictive
experimental limit on the coupling constant of axions to
photons\cite{castprl}.
 The CAST second phase, 2005-2007, will
allow us to scan axion masses greater than 0.02~eV/c$^2$. This is
made possible by filling the magnet bore with a buffer gas ($^4$He
or $^3$He) allowing the photon to acquire an effective mass.
During 2005 a He gas system has been designed and constructed,
allowing systematic changes of the He pressure. Cold polypropylene
windows have been installed in the magnet bore to minimize the
thermal coupling between the cold bore and the outside of the
magnet.

\section{Performance of the Micromegas detector during phase I}
\subsection{Description}
The Micromegas detector is a double gap chamber. It consists of a
conversion gap separated from an amplification gap by a
gauze-light electroformed conducting micromesh. A full description
of the detection principle can be found in Giomataris {\it et
al.}\cite{ioa}. For CAST, the conversion gap is 20 (30)~mm thick
and the amplification gap used was of 50 (100)~$\mu$m in the 2003
(2004) data taking. The gas mixture is 95~\% Argon and 5~\%
Isobutane. The field applied to the amplification gap is about 40
times higher than the conversion field.
The charge is collected on a two dimensional readout by means of X
and Y strips of 350~$\mu$m pitch.The mechanical pieces of the
detector are of Plexiglas because of its low natural radioactivity
to limit as much as possible the inherent background of the
detector.
The Micromegas detector, which operates at atmospheric pressure,
is interfaced with the LHC magnet by two thin vacuum windows.
These two windows, made of polypropylene, act as a buffer for the
pressure gradient and at the same time they need to maximize the
x-ray transmission in the energy range of interest for the axion
search. The integrated x-ray efficiency of the detector in the
solar axion energy spectrum (1-8 keV) is of 73 ~\% taking into
account the loss in transmission due to the thin windows as well
as the conversion efficiency of the gas.

\subsection{Results}
The detector has run successfully during the first phase of the
experiment. The Micromegas detector records tracking data at
sunrise, and during the rest of the day background data is taken.
The detector is calibrated daily using a $^{55}$Fe source. The
energy resolution at 6~keV is 20~\% (FWHM). Tracking and
background data have been analysed. Signal events (photons with a
mean energy of 2-8~keV) have a well defined signature giving a
typical cluster in the read out strips and a typical pulse in the
micromesh. Background events, coming from cosmic rays and natural
radioactivity, give out a bigger cluster in the strips, and the
pulse shape in the micromesh is very different favouring an
efficient rejection based on the micromesh pulse shape and on the
cluster topology.

\begin{figure}
\begin{center}
\centerline{\epsfxsize=6cm \epsfbox{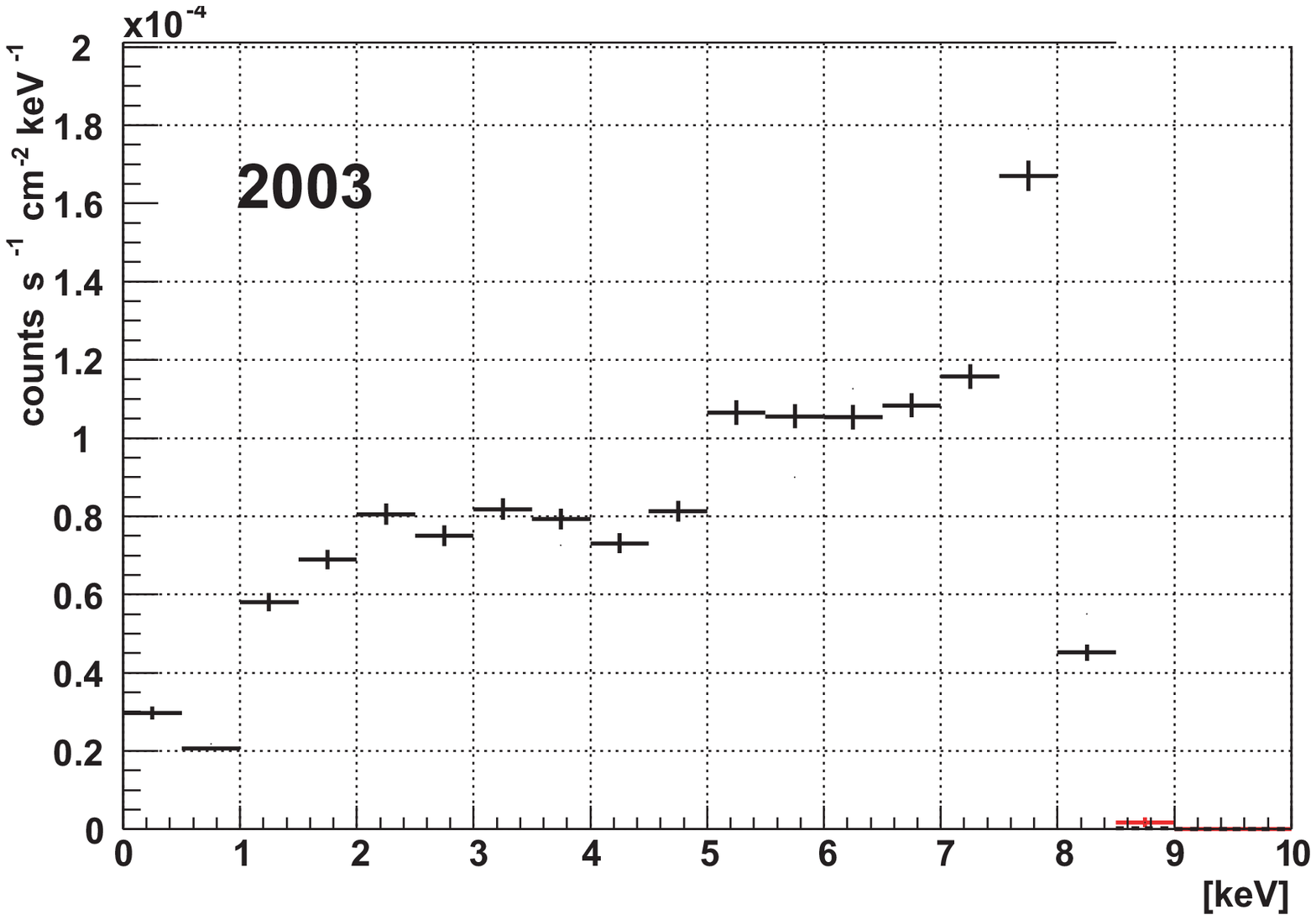} \epsfxsize=6cm
\epsfbox{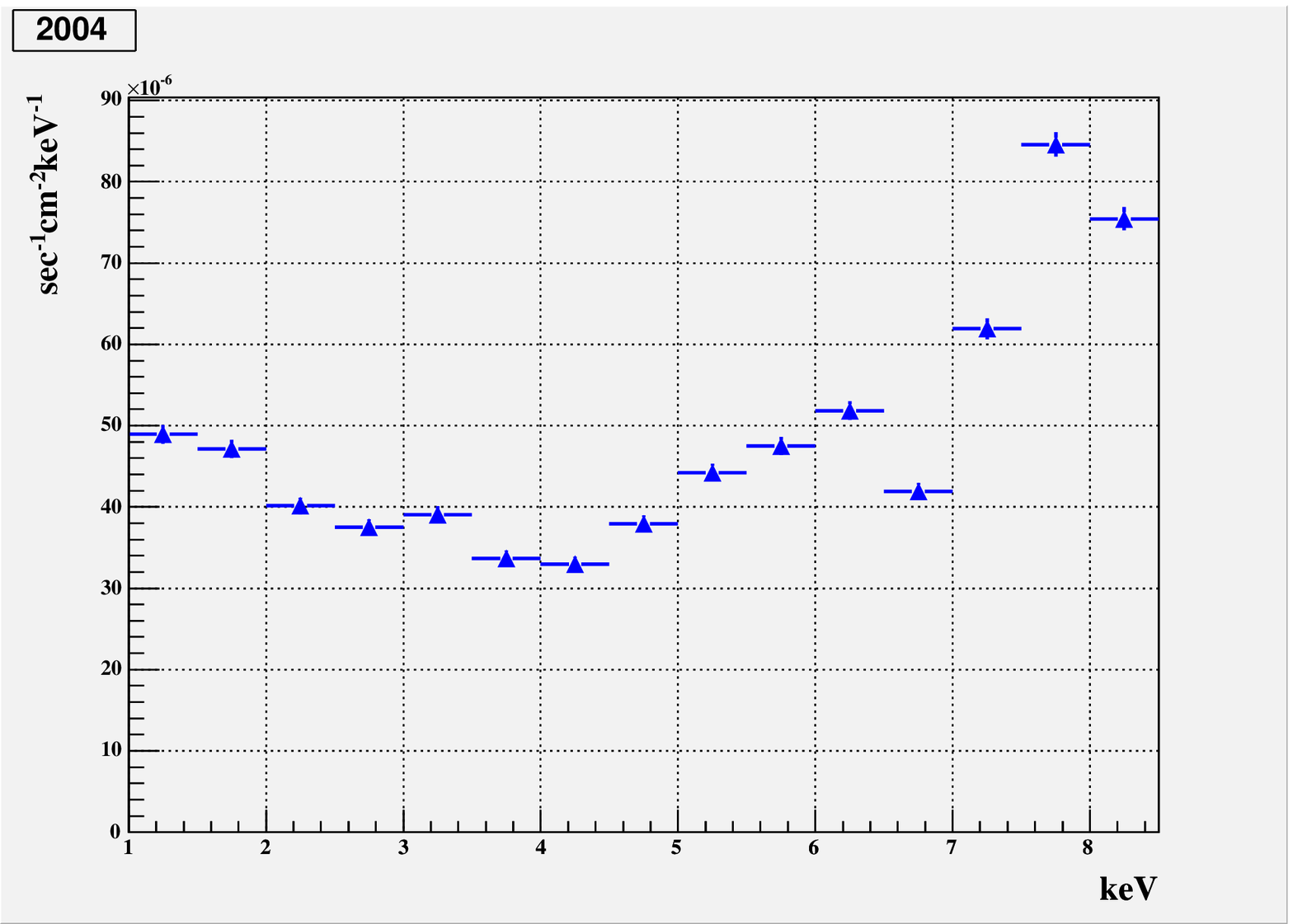}} \caption{Background spectra for the
2003 and 2004 data. \label{fig:bkgspectra}}
\end{center}
\end{figure}

Figure~\ref{fig:bkgspectra} shows the energy spectra for
background events for 2003 and 2004 data. In these plots, a flat
background is observed with a superimposed peak at 8~keV. This
peak has been identified as the copper fluorescence peak coming
from the detector materials (micromesh and readout strips). Thanks
to the upgrading of the detector in 2004 and the elimination of
residual cross-talk between strips, the background rejection was
improved of a factor 3 with respect to 2003, achieving a level of
$5 \times 10^{-5}$ counts keV$^{-1}$cm$^{-2}$s$^{-1}$.
The combined analysis of the three CAST detectors for the 2003
data has resulted in a limit on the coupling constant of photon to
axions\cite{castprl}. This limit, $1.16 \times 10^{-10}$
GeV$^{-1}$ for m$_a<0.02$~eV, improves the existing experimental
limit by a factor 6. The 2004 combined analysis should allow us to
extract a limit close to the theoretical bound coming from
astrophysical considerations.

\section{The new Micromegas line for phase II}
\begin{figure}
\begin{center}
\epsfxsize=10cm   
\epsfbox{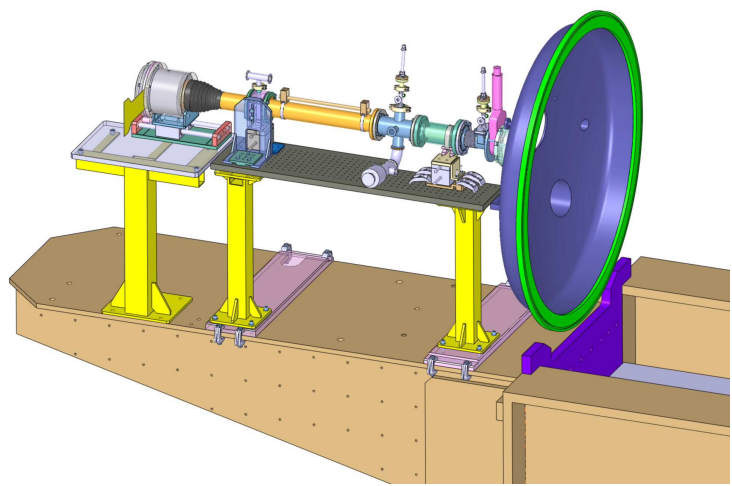}
\caption{The new Micromegas line consisting of a new detector with
integrated x-rays optics and shielding. \label{fig:newline}}
\end{center}
\end{figure}

During the transformation of the magnet for phase II, the
Micromegas group has been designing an upgraded Micromegas
detector surrounded by a shielding and coupled to an x-ray optic,
as shown in  Figure~\ref{fig:newline}. This new line should be
installed at the CAST experiment during spring 2006. These
upgrades will improve significantly the performance of the
detector. First, the x-ray optic, a concentrator with a 1.3~m
focal length and 47~mm diameter, will allow us to increase the
signal to noise ratio by a factor $\sim 100$ by focusing the
photon flux in a 2~mm spot. This concentrator will consist of 14
nested polycarbonate shells, each 125~mm long and coated with
iridium. The optic will transmit $\sim36\%$ of the 0.5-10~keV flux
emerging from the magnet bore. Second, the shielding, composed of
copper, lead, cadmium, nitrogen and polyethylene, is expected to
reduce the background by a factor of 4. Third, by changing the gas
of the chamber from Argon to Xenon, the photon conversion
probability can be improved by at least $\sim 10\%$. The new
detector will be running with the same electronics and acquisition
developed for phase I. First tests of a prototype of the detector
have started. The complete line, with the integrated optics, is
expected to be operational for characterisation tests and
calibration at the PANTER x-ray test facility in Munich beginning
of 2006. Recent developments on the integration of Micromegas with
pixel CMOS readout\cite{medipix}, may lead to an ultimate upgrade
of the detector for the 2007 data taking to profit of the spatial
and energy resolution of pixel sensors.

\section{Conclusions}
A low background Micromegas detector has been operating at the
CAST experiment achieving a remarkable background rejection ($5
\times 10^{-5}$ counts keV$^{-1}$cm$^{-2}$s$^{-1}$). For the CAST
second phase, an upgraded Micromegas detector with integrated
x-ray optics and shielding will be installed at the experiment
during spring 2006. These improvements should lead to a reduction
of the background level by at least two orders of magnitude.
\section*{Acknowledgments}
This work has been performed within the CAST collaboration. We
thank our colleagues for their support. Work at LLNL was performed
under the auspices of the U.S. Department of Energy by the
University of California Lawrence Livermore National Laboratory
under Contract No. W-7405-ENG-48.

\end{document}